\documentclass[prb,onecolumn,12pt]{revtex4-1}
\usepackage{epsfig}
\usepackage{amssymb}
\usepackage{threeparttable}
\usepackage{graphicx}
\usepackage{natbib}
\usepackage{longtable}
\usepackage{txfonts}
\usepackage{color}

\begin{document}
\title{Unexpected phonon-transport properties of stanene among 2D group-IV materials from \textit{ab initio}}
\author{Bo Peng$^1$, Hao Zhang$^{1,*}$, Hezhu Shao$^{2}$, Yuanfeng Xu$^1$, Gang Ni$^1$, Rongjun Zhang$^1$ and Heyuan Zhu$^1$}
\affiliation{$^1$Shanghai Ultra-precision Optical Manufacturing Engineering Research Center and Key Laboratory of Micro and Nano Photonic Structures (Ministry of Education), Department of Optical Science and Engineering, Fudan University, Shanghai 200433, China\\
$^2$Ningbo Institute of Materials Technology and Engineering, Chinese Academy of Sciences, Ningbo 315201, China}

\begin{abstract}
It has been argued that stanene has lowest lattice thermal conductivity among 2D group-IV materials because of largest atomic mass, weakest interatomic bonding, and enhanced ZA phonon scattering due to the breaking of an out-of-plane symmetry selection rule. However, we show that although the lattice thermal conductivity $\kappa$ for graphene, silicene and germanene decreases monotonically with decreasing Debye temperature, unexpected higher $\kappa$ is observed in stanene. By enforcing all the invariance conditions in 2D materials and including Ge $3d$ and Sn $4d$ electrons as valence electrons for germanene and stanene respectively, the lattice dynamics in these materials are accurately described. A large acoustic-optical gap and the bunching of the acoustic phonon branches significantly reduce phonon scattering in stanene, leading to higher thermal conductivity than germanene. The vibrational origin of the acoustic-optical gap can be attributed to the buckled structure. Interestingly, a buckled system has two competing influences on phonon transport: the breaking of the symmetry selection rule leads to reduced thermal conductivity, and the enlarging of the acoustic-optical gap results in enhanced thermal conductivity. The size dependence of thermal conductivity is investigated as well. In nanoribbons, the $\kappa$ of silicene, germanene and stanene is much less sensitive to size effect due to their short intrinsic phonon mean free paths. This work sheds light on the nature of phonon transport in buckled 2D materials.
\end{abstract}

\maketitle

\section{INTRODUCTION}
Two-dimensional (2D) materials are one of the most active areas of nanomaterials research due to their potential for integration into next-generation electronic and energy conversion devices \cite{Ferrari2015,Novoselov2012,Klinovaja2013}. Graphene, the most widely studied 2D material, is predicted to possess massless Dirac fermions, where the Fermi velocity $v_F$ of the graphene substitutes for the speed of light \cite{Geim2009,Zhang2005,Issi2014,Luican-Mayer2014}. Recently, the other 2D group-IV materials, silicene, germanene and stanene, have been realized by epitaxial growth on substrates \cite{Vogt2012,Davila2014,Zhu2015}, and attracted tremendous interest due to their extraordinary properties. Unlike graphene, silicene, germanene and stanene have buckled honeycomb structure and large spin-orbital coupling (SOC) strength, which opens a nontrivial band gap at the Dirac point, resulting in significant quantum spin Hall (QSH) effect \cite{Liu2011a,Liu2011,Xu2013,Matusalem2015}. Such properties provide opportunities for spintronic applications. Furthermore, the superior mechanical flexibility of silicene compared to graphene makes it highly adaptable for flexible nanoelectronics \cite{Balendhran2015}. In addition, silicene and germanene are expected to be easily incorporated into the silicon-based microelectronics industry. Apart from silicene and germanene, recent study has found that stanene can provide enhanced thermoelectricity \cite{Xu2014a}.

Thermal transport plays an important role in these applications. With extremely high thermal conductivity, graphene has great potential in applications including electronic cooling \cite{Balandin2008}; while for application in thermoelectric (TE) energy conversion, it is important to reduce the lattice thermal conductivity of a material while maintaining a high electrical conductivity, which often conflict with each other \cite{Qin2014,Fan2012}. Silicene, germanene and stanene are expected to be topological insulators \cite{Ezawa2015,Matusalem2015,Liu2011a,Liu2011,Xu2013}, and the TE figure of merit $zT$ can be improved by optimizing the geometry size to decrease the lattice thermal conductivity and maximize the contribution of the gapless edge states to the electron transport \cite{Xu2014a}. Thus, systematic investigation of phonon transport properties for 2D group-IV materials is needed. 

Detailed theoretical investigations have predicted that the thermal conductivity $\kappa$ of graphene and silicene are in the range of 2000-5000 W/mK and 20-30 W/mK, respectively \cite{Nika2009,Lindsay2010,Li2012b,Pei2013,Hu2013,Issi2014,Fugallo2014,Lindsay2014,Cepellotti2015,Gu2015,Xie2016}. Moreover, strain effects on lattice thermal conductivity of 2D group-IV crystals have been investigated \cite{Hu2013,Xie2016,Kuang2016}. However, due to the violation of crystal symmetry, translational invariance and rotational invariance in 2D materials in the computational algorithms \cite{Carrete2016}, in silicene, germanene and stanene, the flexural acoustic branch usually has a linear component \cite{Kuang2016,Nissimagoudar2016}, which significantly influence the phonon transport. To get a more precise estimation, in our calculations, all the invariance conditions in 2D materials such as translations, rotations, and crystal symmetry are enforced \cite{Carrete2016}. In addition, some previous calculations even predict an instability for germanene and stanene with a small region of negative frequencies near the $\Gamma$ point \cite{Kuang2016}. Here we find that treating Ge $3d$ and Sn $4d$ electrons as valence electrons is required to accurately describe the lattice dynamics of germanene and stanene, which is similar to the case in InP \cite{Mukhopadhyay2014}.

In comparison to studies that focus on only one material, the general nature of phonon transport properties in all these 2D group-IV materials is less investigated, and a comprehensive understanding is still lacking. Traditionally there are four factors that determine the lattice thermal conductivity, including (i) average atomic mass, (ii) interatomic bonding, (iii) crystal structure, and (iv) anharmonicity \cite{Slack1973321}. According to the Slack's theory, low average atomic mass and strong interatomic bonding imply a high Debye temperature, which leads to a high thermal conductivity. This has been observed in monolayer transition metal dichalcogenides MX$_2$ (M=Mo,W; X=S,Se) in our previous work \cite{Peng2016a}. However, recent studies have found that, phonon vibrational properties such as acoustic-optical (a-o) gap and acoustic bunching also have significant influence on $\kappa$ \cite{Lindsay2013,Lindsay2013a,Broido2013,Jain2014,Gu2014}, leading to unexpected phonon transport properties. It has been reported that in silicene and germanene, the a-o gap enlarges with increasing buckling height \cite{Huang2015a}. Thus it is also interesting to examine that if unexpected phonon transport behavior exists in 2D group-IV materials.

\begin{figure*}
\centering
\includegraphics[width=0.9\linewidth]{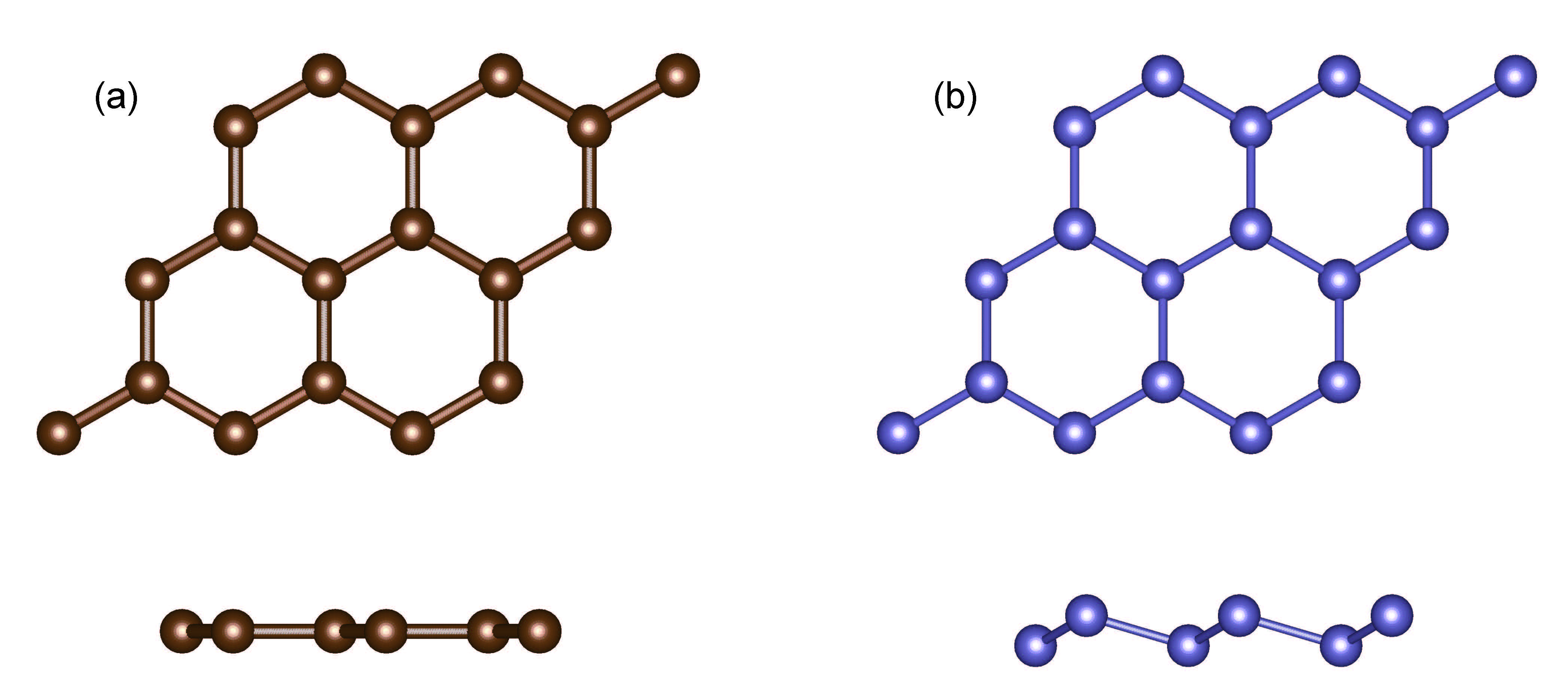}
\caption{Top view and side view of (a) planar hexagonal crystal structure of graphene, and (b) buckled hexagonal crystal structure of silicene, germanene, and stanene.}
\label{structure} 
\end{figure*}

In this paper, we investigate the lattice thermal conductivity $\kappa$ of 2D group-IV materials using first-principles calculations and an iterative solution of the Boltzmann transport equation (BTE) for phonons \cite{ShengBTE,Li2012,Li2012a}. In contrast to the Slack's theory, unexpected higher lattice thermal conductivity in stanene is obtained though the Debye temperature of stanene is nearly two times lower than that of germanene. Although it has been demonstrated recently that buckled structures usually lead to lower thermal conductivity due to the breaking of the out-of-plane symmetry \cite{Lindsay2010,Jain2015}, here we show that these buckled structures also result in a large a-o gap, which tends to reduce the thermal resistance. Therefore to estimate accurately the thermal transport in 2D group-IV materials, detailed analysis of the scattering mechanism is required to understand the competing effects between conventional Slack’s theory and certain phonon vibrational properties such as the a-o gap. The size dependence of lattice thermal conductivity is investigated as well for the purpose of the design of TE nanostructures.

\section{METHODOLOGY}

The in-plane $\kappa$ is isotropic and can be calculated as a sum of contribution of all the phonon modes $\lambda$ \cite{Li2013a,Ouyang2015}, which comprises both a phonon branch index $p$ and a wave vector $\textbf{q}$,
\begin{equation}\label{kappa-eq}
\kappa=\kappa_{\alpha\alpha}=\frac{1}{V}\sum_{\lambda}C_{\lambda}v_{\lambda\alpha}^2\tau_{\lambda\alpha},
\end{equation}
where $V$ is the crystal volume, $C_{\lambda}$ is the heat capacity per mode, $v_{\lambda\alpha}$ and $\tau_{\lambda\alpha}$ are the group velocity and relaxation time of mode $\lambda$ along $\alpha$ direction, respectively. We use the nominal layer thicknesses $h$ = 3.35 \AA, 4.20 \AA, 4.22 \AA\ and 4.34 \AA\ for graphene, silicene, germanene and stanene, corresponding to the van der Waals radii of carbon, silicon, germanium and tin atoms, respectively \cite{Hu2013,Kuang2016}. The lattice thermal conductivity can be calculated iteratively using the ShengBTE code \cite{Omini1996,ShengBTE,Li2012a,Li2012}. The only inputs are harmonic and anharmonic interatomic force constants (IFCs), which are obtained from first-principles calculations.

All the calculations are performed using the Vienna \textit{ab-initio} simulation package (VASP) based on density functional theory (DFT) \cite{Kresse1996}. We choose the generalized gradient approximation (GGA) in the Perdew-Burke-Ernzerhof (PBE) parametrization for the exchange-correlation functional. A plane-wave basis set is employed with kinetic energy cutoff of 600 eV. For germanene and stanene, we use the projector-augmented-wave (PAW) potential with 3\textit{d} electrons of Ge and 4\textit{d} electrons of Sn described as valence, respectively. A 21$\times$21$\times$1 \textbf{k}-mesh is used during structural relaxation for the unit cell until the energy differences are converged within 10$^{-6}$ eV, with a Hellman-Feynman force convergence threshold of 10$^{-4}$ eV/\AA. We maintain the interlayer vacuum spacing larger than 15 \AA\ to eliminate interactions between adjacent supercells. 

The harmonic IFCs are obtained by density functional perturbation theory (DFPT) using the supercell approach, which calculates the dynamical matrix through the linear response of electron density \cite{DFPT}. A 5$\times$5$\times$1 supercell with 5$\times$5$\times$1 \textbf{q}-mesh is used. The harmonic IFCs are symmetrized by enforcing all the invariance conditions in 2D materials \cite{Carrete2016}. Using the harmonic IFCs, the phonon dispersion relation can be obtained, which determines the group velocity $v_{\lambda\alpha}$ and specific heat $C_{\lambda}$. The third-order anharmonic IFCs play an important role in calculating the three-phonon scattering rate, which is the inverse of $\tau_{\lambda\alpha}$.  The anharmonic IFCs are calculated using a supercell-based, finite-difference method \cite{Li2012}, and the same 5$\times$5$\times$1 supercell with 5$\times$5$\times$1 \textbf{q}-mesh is used. We include the interactions with the eighth nearest-neighbor atoms for graphene (6.0 \AA), silicene (9.5 \AA), germanene (9.8 \AA), and stanene (11.3 \AA).

The convergence of thermal conductivity with respect to $\textbf{q}$ points is tested in our calculation. A discretizationa of the Brillouin zone (BZ) into a $\Gamma$-centered regular grid of 200$\times$200$\times$1 $\textbf{q}$ points is introduced. In order to enforce the conservation of energy in the three-phonon processes, a Gaussian function is used with scale parameter for broadening chosen as 1.0 \cite{Li2012a}.

\begin{table*}
\centering
\caption{Calculated lattice constant $a$, buckling height $h$, nearest-neighbor distance $d$, and angle between neighboring bonds $\theta$ of all studied 2D group-IV crystals. Other theoretical data are also listed in parentheses for comparison.}
\begin{tabular}{ccccc}
\hline
 & $a$ (\AA) & $h$ (\AA) & $d$ (\AA) & $\theta$ ($^\circ$) \\
\hline
graphene & 2.47 (2.46 \cite{Sahin2009}) & - & 1.43 (1.42 \cite{Sahin2009}) & 120.0 (120.0 \cite{Sahin2009}) \\
silicene & 3.87 (3.87 \cite{Scalise2013}) & 0.44 (0.44 \cite{Scalise2013}) & 2.28 (2.28 \cite{Scalise2013}) & 116.3 (116.4 \cite{Sahin2009}) \\
germanene & 4.06 (4.07 \cite{Scalise2013}) & 0.69 (0.70 \cite{Scalise2013}) & 2.44 (2.44 \cite{Scalise2013}) & 112.3 (113.0 \cite{Sahin2009}) \\
stanene & 4.67 (4.67 \cite{Matusalem2015}) & 0.85 (0.85 \cite{Matusalem2015}) & 2.83 (2.83 \cite{Cherukara2016}) & 111.3 (109 \cite{Mojumder2015}) \\
\hline
\end{tabular}
\label{lattice}
\end{table*}

\section{RESULTS AND DISCUSSION}
\subsection{Crystal structures and Phonon dispersions}

\begin{figure*}
\centering
\includegraphics[width=0.9\linewidth]{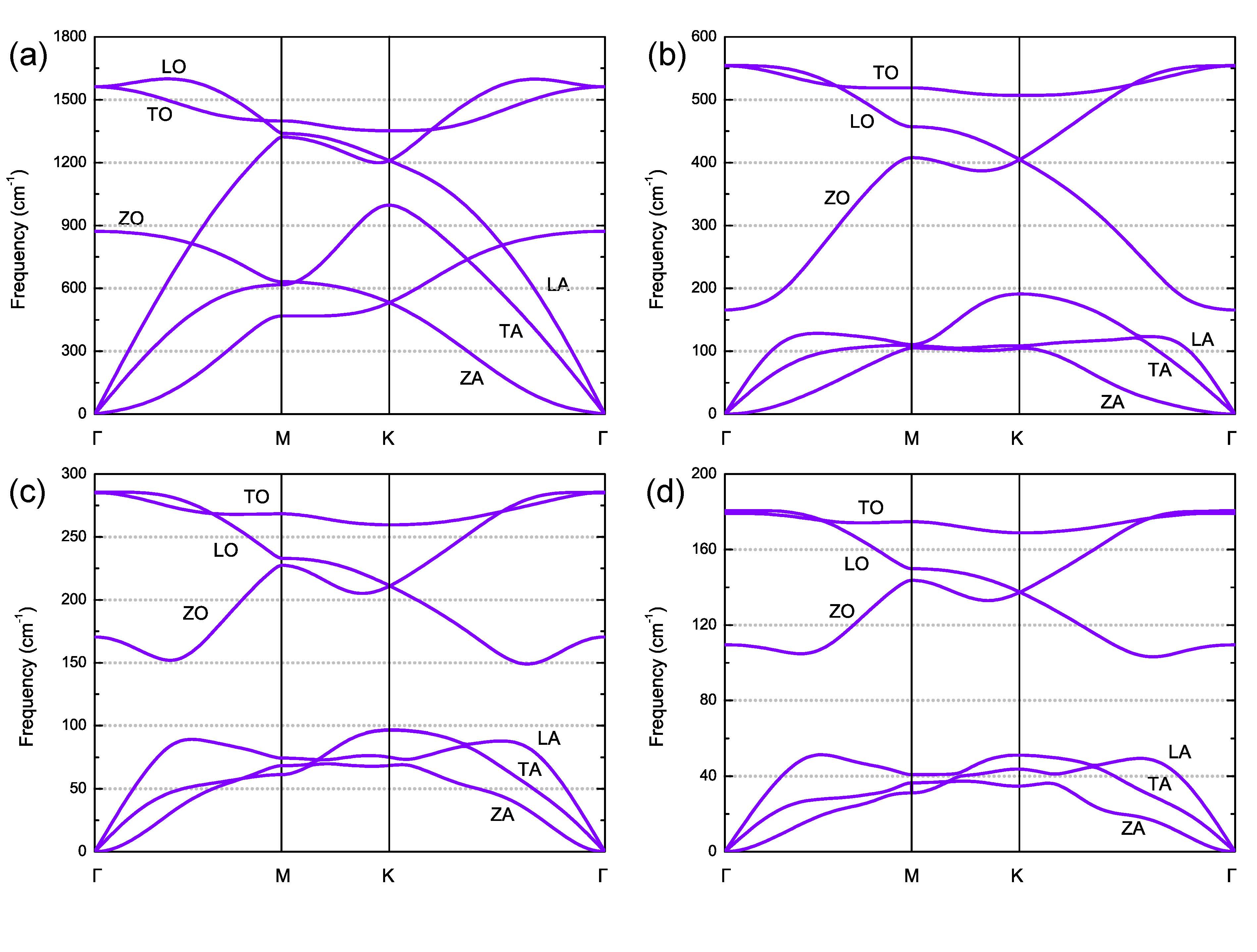}
\caption{Phonon dispersion for (a) graphene, (b) silicene, (c) germanene and (d) stanene along different symmetry lines. Note the different scales along the vertical axis.}
\label{phonon-dispersion} 
\end{figure*}

Fig.~\ref{structure} shows the optimized structure of 2D group-IV crystals. A low-buckled configuration is found in silicene, germanene and stanene in Fig.~\ref{structure}(b), which is different from the planar geometry of graphene in Fig.~\ref{structure}(a). The optimized geometries of all studied 2D group-IV crystals are listed in Table~\ref{lattice}, which are in good agreement with previous works \cite{Sahin2009,Matusalem2015,Scalise2013}. There are two atoms in the unit cell, corresponding to three acoustic and three optical phonon branches.

The phonon dispersion determines the allowed three-phonon scattering processes, and plays a significant role in precise calculation of phonon transport properties. Fig.~\ref{phonon-dispersion} shows the calculated phonon dispersions of all studied 2D group-IV crystals, which is in agreement with previous works \cite{Nika2009,Scalise2013}. Similar to other 2D materials \cite{Molina-Sanchez2011,Qin2015}, the longitudinal acoustic (LA) and transverse acoustic (TA) branches of group-IV materials are linear near the $\Gamma$ point, while the z-direction acoustic (ZA) branch is quadratic near the $\Gamma$ point. For strict 2D, even in a buckled system, the ZA branch of unstrained monolayer material should be quadratic in the long-wavelength limit \cite{Carrete2016}. However, in previous work, linear dispersion of the ZA branch is observed \cite{Kuang2016,Nissimagoudar2016}, which significantly influence the phonon transport. In particular, some even report that the ZA dispersion of germanene and stanene shows imaginary frequencies \cite{Tang2014,Kuang2016}, which is possibly due to improper pseudopotential used in these works without considering the role of Ge $3d$ and Sn $4d$ electrons in accurate describing lattice dynamics \cite{Mukhopadhyay2014} (for details, see Fig.~S1 in the supplemental material).

\begin{table*}
\centering
\caption{The average atomic mass $\overline{M}$, highest phonon frequency $\omega_m$, Young's modulus $Y_{2D}$, sound velocities $v_s$ and Debye temperature $\Theta_D$ of all studied 2D group-IV crystals.}
\begin{tabular}{ccccccc}
\hline
 & $\overline{M}$ (amu) & $Y_{2D}$ (N/m) & $\omega_m$ (cm$^{-1}$) & $v_s$(TA) (km/s) & $v_s$(LA) (km/s) & $\Theta_D$ (K) \\
\hline
graphene & 12.01 & 341.9 & 1599 & 13.85 & 21.59 & 2539 \\
silicene & 28.09 & 62.8 & 554 & 5.77 & 9.82 & 680 \\
germanene & 72.64 & 43.5 & 286 & 3.12 & 5.32 & 352 \\
stanene & 118.71 & 23.5 & 182 & 2.02 & 3.65 & 198 \\
\hline
\end{tabular}
\label{phonon-debye}
\end{table*}

\begin{figure*}[h]
\centering
\includegraphics[width=0.9\linewidth]{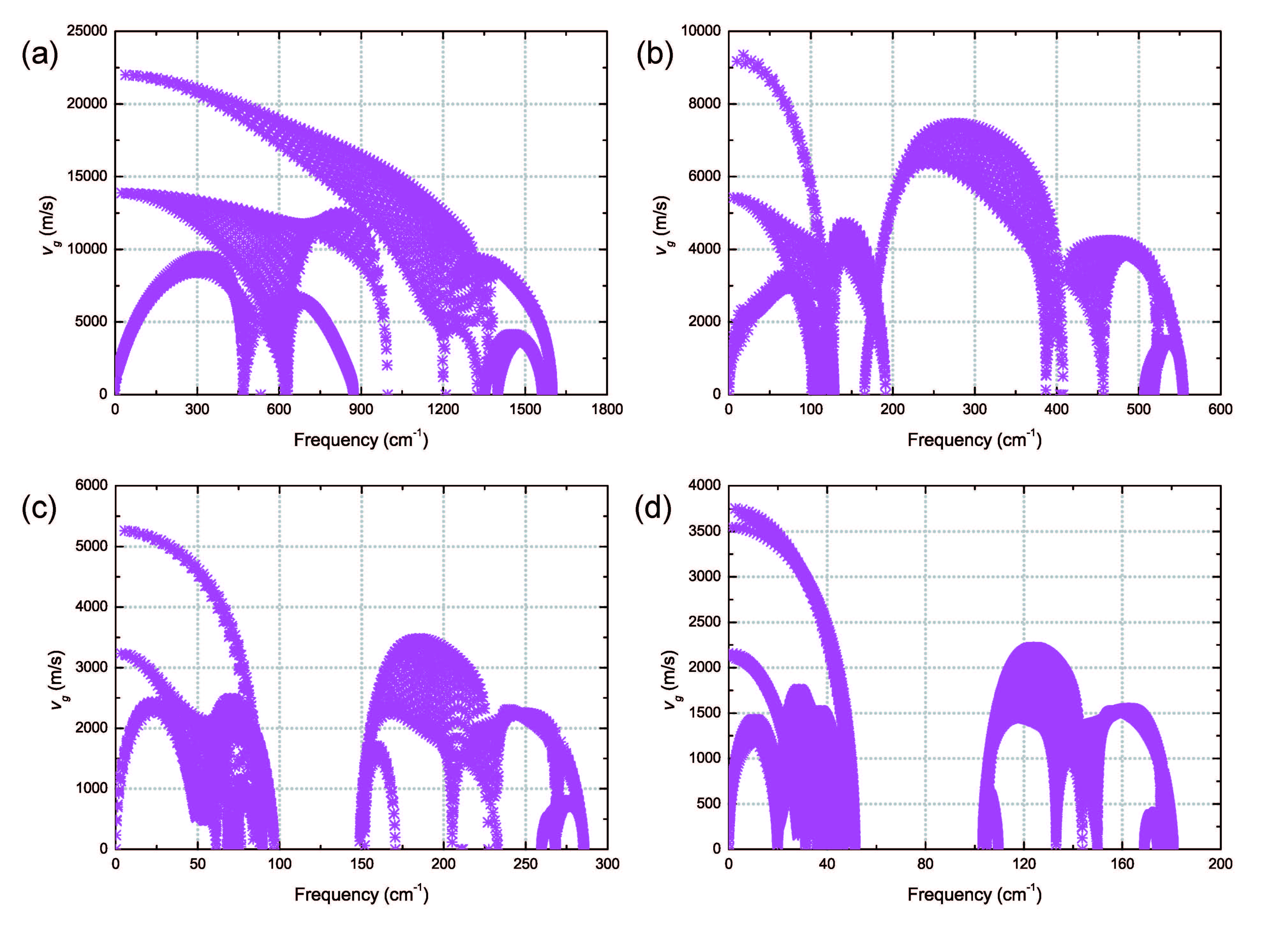}
\caption{Calculated phonon group velocities for (a) graphene, (b) silicene, (c) germanene and (d) stanene in the irreducible BZ.}
\label{vg} 
\end{figure*}

Similar to the monatomic linear chain model, the scale of the phonon branch is dominated by the average atomic mass and interatomic bonding. The bond strength is estimated by the 2D Young's modulus, which is calculated from the elastic tensor coefficients with ionic relaxations using the finite differences method \cite{LePage2002,Wu2005}. As shown in Table~\ref{phonon-debye}, the phonon frequency decreases with increasing mass and decreasing bond strength. Stiffness is resistance to elastic deformation and can be used to measure the sound velocity in crystals \cite{Tohei2006,Politano2015},
\begin{equation}
v_s(\textsc{LA})=\sqrt{\frac{B+G}{\rho_{2D}}},
\end{equation}
\begin{equation}
v_s(\textsc{TA})=\sqrt{\frac{G}{\rho_{2D}}},
\end{equation}
where $B$ is the in-plane stiffness and $G$ is the shear modulus, given by
\begin{equation}
B=\frac{Y_{2D}}{2(1-\nu)},
\end{equation}
\begin{equation}
G=\frac{Y_{2D}}{2(1+\nu)},
\end{equation}
where $\nu$ is the Poisson's ratio (for details, see Table~S1 in the supplemental material). The calculated sound velocities of the TA and LA branches are in good agreement with the phonon group velocities for the TA and LA modes in long-wavelength limit, as shown in Fig.~\ref{vg}.

The Debye temperature $\Theta_D$ can be calculated from the average sound velocity $v_s$ \cite{Shao2016}
\begin{equation}
\label{Debye-stiffness}
\Theta_D=\frac{\hbar v_s}{k_B} \bigg( \frac{4\pi N}{S} \bigg)^{1/2},
\end{equation}
where $\hbar$ is the reduced Planck constant, $k_B$ is the Boltzmann constant, $N$ is the number of atoms in the cell, $S$ is the area of the unit cell, and $v_s$ is the average sound velocity given by 
\begin{equation}
v_s=\bigg[\frac{1}{3} \big( \frac{1}{v_l^3} + \frac{2}{v_t^3} \big) \bigg]^{-1/3},
\end{equation}
As shown in Table~\ref{phonon-debye}, $\Theta_D$ decreases monotonically from C to Si to Ge and to Sn. The Debye temperature is a measure of the temperature above which all modes begin to be excited and below which modes begin to be frozen out \cite{Nakashima1992}. Traditionally, low $\Theta_D$ indicates low thermal condctivity according to the Slack's theory \cite{Slack1973321,Peng2016a}. That is because lower Debye temperature indicates lower phonon velocities and lower acoustic-phonon frequencies, and the latter tend to increase phonon populations, which subsequently increase phonon scattering rates \cite{Nakashima1992,Lindsay2013,Peng2016a}.

\begin{figure*}[h]
\centering
\includegraphics[width=0.9\linewidth]{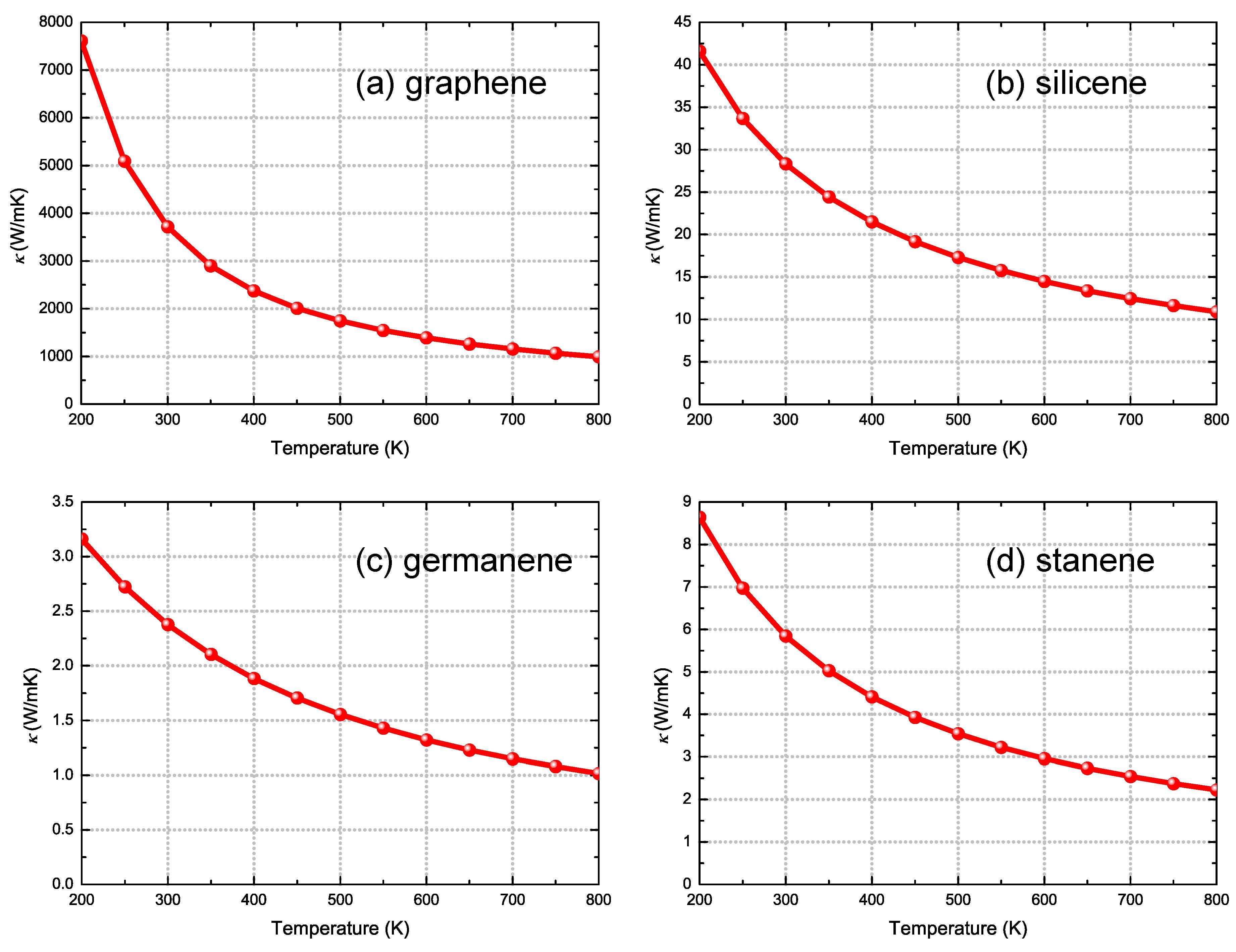}
\caption{Thermal conductivity as a function of temperature for (a) graphene, and (b) silicene, (c) germanene and (d) stanene at natural isotopic abundances.}
\label{kappa} 
\end{figure*}

\subsection{Thermal conductivity}

Fig.~\ref{kappa} presents the lattice thermal conductivity for all four materials with naturally occurring isotope concentrations. The thermal conductivity decreases with increasing temperature, as
expected for a phonon-dominated crystalline material. For naturally occurring graphene, the $\kappa_{nat}$ is 3716.6 W/mK, while for isotopically pure graphene, the intrinsic $\kappa_{pure}$ is 4242.3 W/mK, which agrees well with the previous values \cite{Balandin2008,Lindsay2010,Seol2010,Chen2012a,Lindsay2014}. The calculated thermal conductivity of silicene is also in good agreement with other theoretical results \cite{Li2012b,Pei2013,Gu2015,Xie2016,Peng2016e}. In previous calculations for germanene and stanene, linear dispersion or even imaginary frequency in the ZA branch is observed, and the former tends to overestimate the lattice thermal conductivity, while the latter tends underestimate it \cite{Nissimagoudar2016,Kuang2016}. It should be noted that, by choosing suitable pseudopotential and enforcing all the invariance conditions, our results for stanene give reasonable predictions, and agree well with previous equilibrium molecular dynamics simulations \cite{Cherukara2016}.

The room-temperature (RT) $\kappa_{nat}$ of all studied 2D group-IV material is listed in Table~\ref{thermal}. In consistent with conventional Slack's theory, from graphene to silicene and to germanene, $\kappa_{nat}$ decreases monotonically with decreasing $\Theta_D$ . However, although the $\Theta_D$ of stanene is nearly two times lower than that of germanene, the RT $\kappa_{nat}$ of stanene is more than two times higher than that of germanene. This unusual behavior is not reflected in the Debye temperature.

We first examine the role of isotopes, since considering isotope abundance, the wide spread of isotopes in the naturally occurring Ge (20.5\% Ge$^{70}$, 27.4\% Ge$^{72}$, 7.8\% Ge$^{73}$, 36.5\% Ge$^{74}$, and 7.8\% Ge$^{76}$) may lead to strong isotopic scattering, and subsequently low lattice thermal conductivity. However, as shown in Table~\ref{thermal}, for silicene, germanene and stanene, the isotopic influence on thermal conductivity is much less obvious than graphene. This difference is due to a simple fact that Si, Ge and Sn atoms are heavier than C. According to Klemens' theory \cite{Klemens1951,Klemens1955}, the isotopic scattering rate depends on the mass ratio $\Delta M/M$, where $\Delta M$ is the mass difference. The larger atomic mass leads to a much lower mass ratio $\Delta M/M$. Thus the isotopic scattering cannot explain unexpected phonon transport behavior in stanene.

\begin{table*}
\centering
\caption{$\kappa_{nat}$, $\kappa_{pure}$ and contribution of different phonon modes branches (ZA, TA, LA, and all optical phonons) towards the $\kappa_{nat}$ in all studied 2D group-IV crystals at 300 K.}
\begin{tabular}{ccccccc}
\hline
 & $\kappa_{nat}$ (W/mK) & $\kappa_{pure}$ (W/mK) & ZA (\%) & TA (\%) & LA (\%) & Optical (\%) \\
\hline
graphene & 3716.6 & 4242.3 & 76.4 & 14.7 & 7.9 & 1.0 \\
silicene & 28.3 & 28.6 & 67.3 & 7.9 & 12.9 & 11.9 \\
germanene & 2.4 & 2.7 & 33.5 & 36.7 & 21.8 & 8.0 \\
stanene & 5.8 & 6.0 & 29.7 & 26.7 & 28.5 & 15.1 \\
\hline
\end{tabular}
\label{thermal}
\end{table*}

To obtain more insight into the unconventional phonon transport properties in stanene, we further compare its specific heat and average values of group velocity with germanene. The specific heat of germanene (49.47 J/K$mol$) is higher than that of stanene (48.71 J/K$mol$), and the average group velocity of germanene (1.39 km/s) is higher than that of stanene (0.89 km/s) as well, which originates from larger slope of phonon branches in the phonon spectrum. 

If not specific heat and group velocity, the differences in phonon scattering processes must be the governing factor according to Eq.~(\ref{kappa-eq}). The phonon scattering depends on two factors \cite{Wu2016}: (i) the strength of each scattering channel, which depends on the anharmonicity of a phonon mode, and is described by the Gr\"uneisen parameter; (ii) the number of channels available for a phonon to get scattered, which is determined by whether there exist three phonon groups that can satisfy both energy and quasi-momentum conservations, and can be characterized by the phase space for three-phonon processes \cite{Lindsay2008,ShengBTE,Peng2016b}.

\begin{figure*}
\centering
\includegraphics[width=0.9\linewidth]{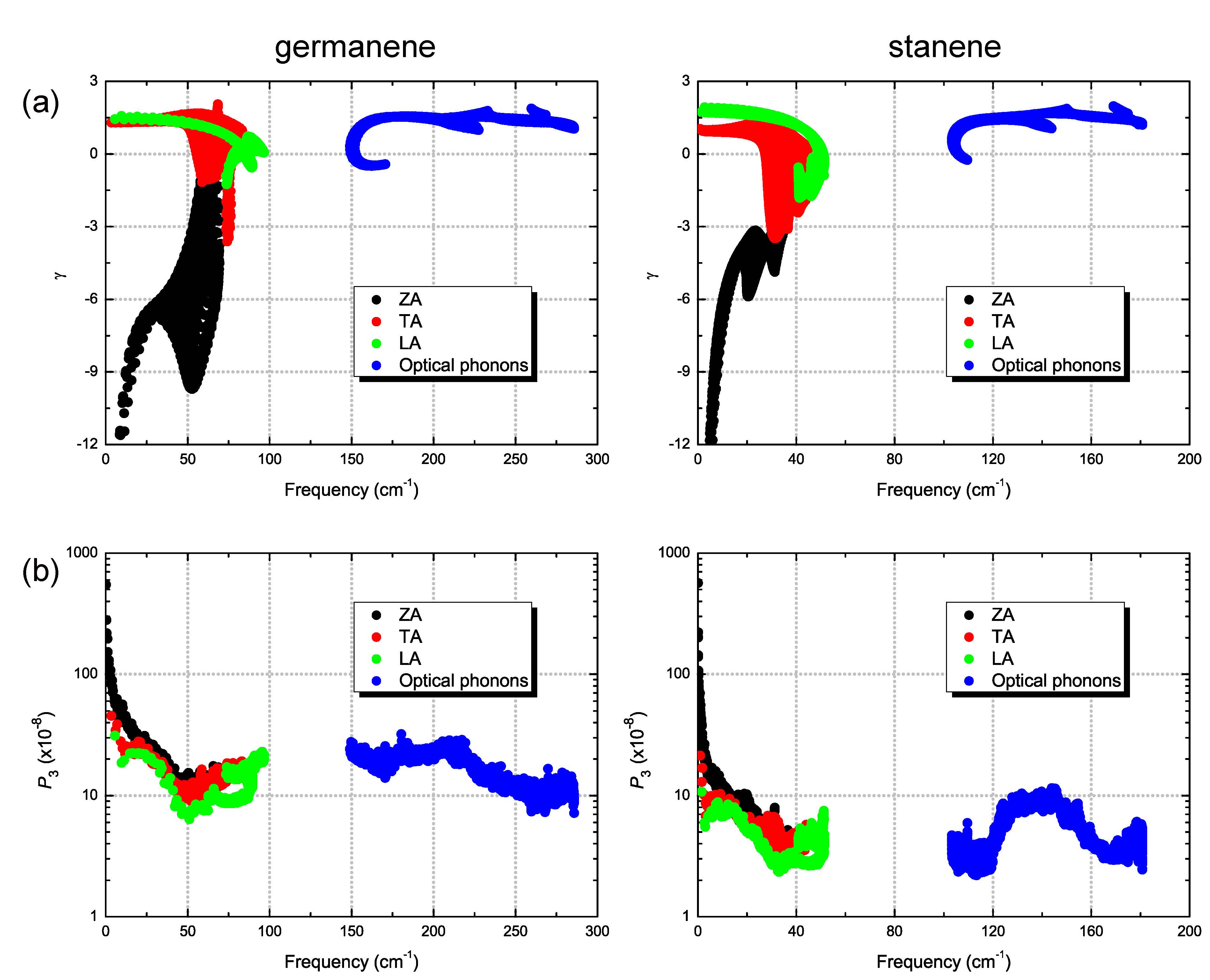}
\caption{Calculated (a) mode Gr\"uneisen parameters $\gamma$ and (b) total phase space for three-phonon processes $P_3$ for germanene and stanene at 300 K.}
\label{scattering} 
\end{figure*}

The mode Gr\"uneisen parameters $\gamma$ provide information about anharmonic interactions, and larger $\gamma$ implies strong anharmonicity, leading to low lattice thermal conductivity. The mode Gr\"uneisen parameters of germanene and stanene are comparable with each other in Fig.~\ref{scattering}(a). Thus the reduced phonon scattering in stanene can be only attibuted to the number of scattering channels. The total phase space for three-phonon processes $P_3$ provides the number of scattering channels available for three-phonon processes of each phonon mode, and larger $P_3$ means low intrinsic lattice thermal conductivity \cite{Ziman,Lindsay2008}. As shown in Fig.~\ref{scattering}(b), a clear decrease in $P_3$ of stanene is observed, indicating fewer scattering channels and subsequently higher lattice thermal conductivity.

To investigate the underlying scattering mechanisms, we extract the contributions of different phonon branches to $\kappa_{nat}$ in Table~\ref{thermal}. The ZA contribution decreases from graphene to silicene to germanene and to stanene, while the LA contribution increases. For acoustic phonons, there are two scattering processes: (i) scattering between three acoustic phonons ($a+a\leftrightarrow a$), and (ii) absorption of two acoustic phonons into one optical phonon or vice versa ($a+a\leftrightarrow o$).

The decreaing ZA contribution is related to the symmetry of crystal structure. For $a+a\leftrightarrow a$ processes, it has been reported that in graphene there is a symmetry selection rule due to the one-atom-thick plane, which strongly restricts anharmonic phonon-phonon scattering of the ZA phonons \cite{Lindsay2010,Lindsay2014}. As a result, only even numbers of ZA phonons can be involved in three-phonon scattering processes (ZA+ZA$\leftrightarrow$TA/LA or ZA+TA/LA$\leftrightarrow$ZA). A buckled system such as silicene, germanene and stanene breaks out the out-of-plane symmetry, and the selection rule does not apply any more, which is similar to blue phosphorene \cite{Jain2015}. Thus the contribution of ZA phonons to $\kappa_{nat}$ decreases with increasing buckling height from silicene to germanene and to stanene.

\begin{figure*}
\centering
\includegraphics[width=0.9\linewidth]{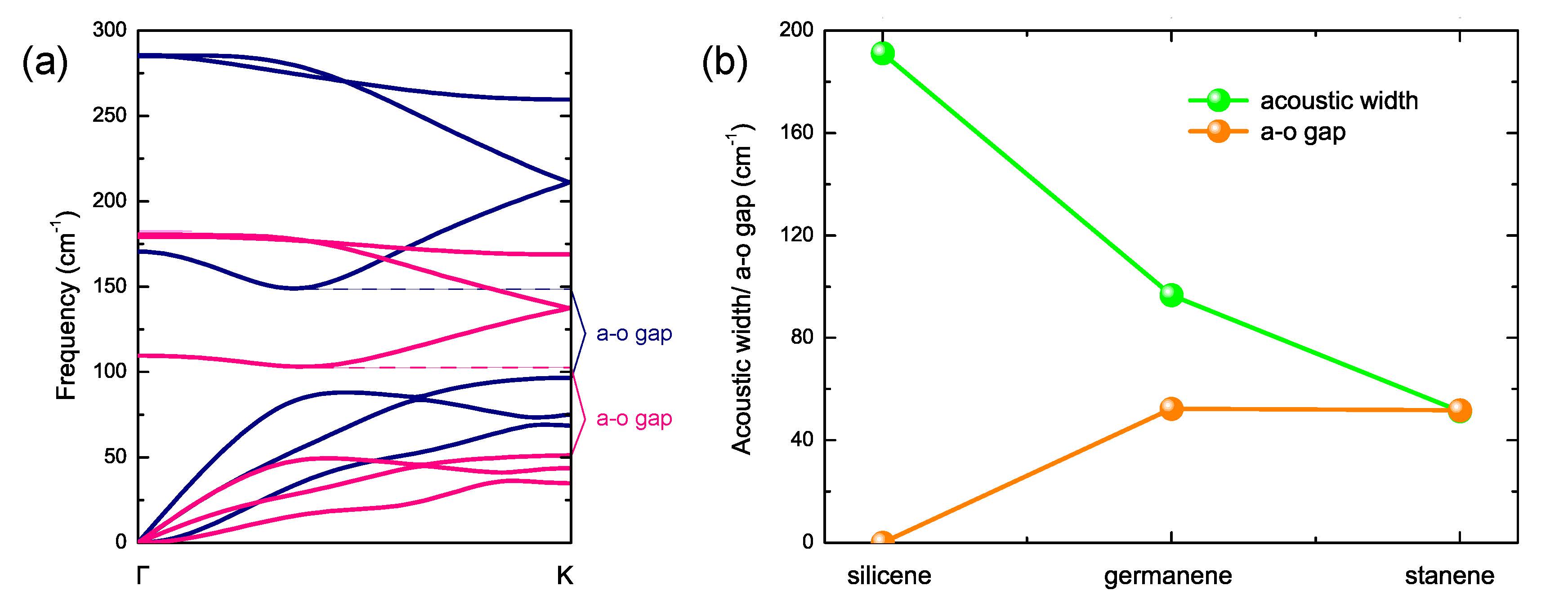}
\caption{(a) Phonon frequency along the high symmetry direction $\Gamma$-K for germanene (blue) and stanene (red). Stanene has a large a-o gap, and the acoustic branches are bunched closer to one another in stanene than in germanene. (b) The acoustic width, a-o gap, and acoustic bunching of silicene, germanene and stanene.}
\label{gap-bunching} 
\end{figure*}

The increasing contribution from LA phonons can be attributed to the a-o gap. For $a+a\leftrightarrow o$ processes, the optical phonons provide important scattering channels for the acoustic phonons. However, $a+a\leftrightarrow o$ scattering channels can be weakened in materials with large a-o gap \cite{Lindsay2012,Lindsay2013,Broido2013,Jain2014}. As shown in Fig.~\ref{gap-bunching}, the acoustic width decreases from silicene to germanene and to stanene, while the a-o gap enlarges, and finally in stanene, the acoustic width becomes smaller than the a-o gap. Thus all optical branches in stanene have frequencies at least twice the highest acoustic branch frequency, and the $a+a\leftrightarrow o$ processes are prohibited by the conservation of energy. The scattering channels for high-frequency LA phonons are suppressed, leading to large LA contribution and reduced thermal resistance. 

In addition, the $a+a\leftrightarrow a$ scattering becomes weak in stanene when high-frequency acoustic branches are bunched together \cite{Lindsay2013,Lindsay2013a,Broido2013,Jain2014}, as shown in Fig.~\ref{gap-bunching}(a). There is another selection rule for $a+a\leftrightarrow a$ processes that prohibits the lowest acoustic phonon branch emitting two or more phonons by anharmonic processes of any order \cite{Ziman,Lax1981}, and subsequently coincident acoustic branches cannot satisfy momentum and energy conservation simultaneously \cite{Lindsay2013,Broido2013}. Thus the phonon scattering between acoustic phonons becomes weaker.

\begin{figure*}
\centering
\includegraphics[width=0.9\linewidth]{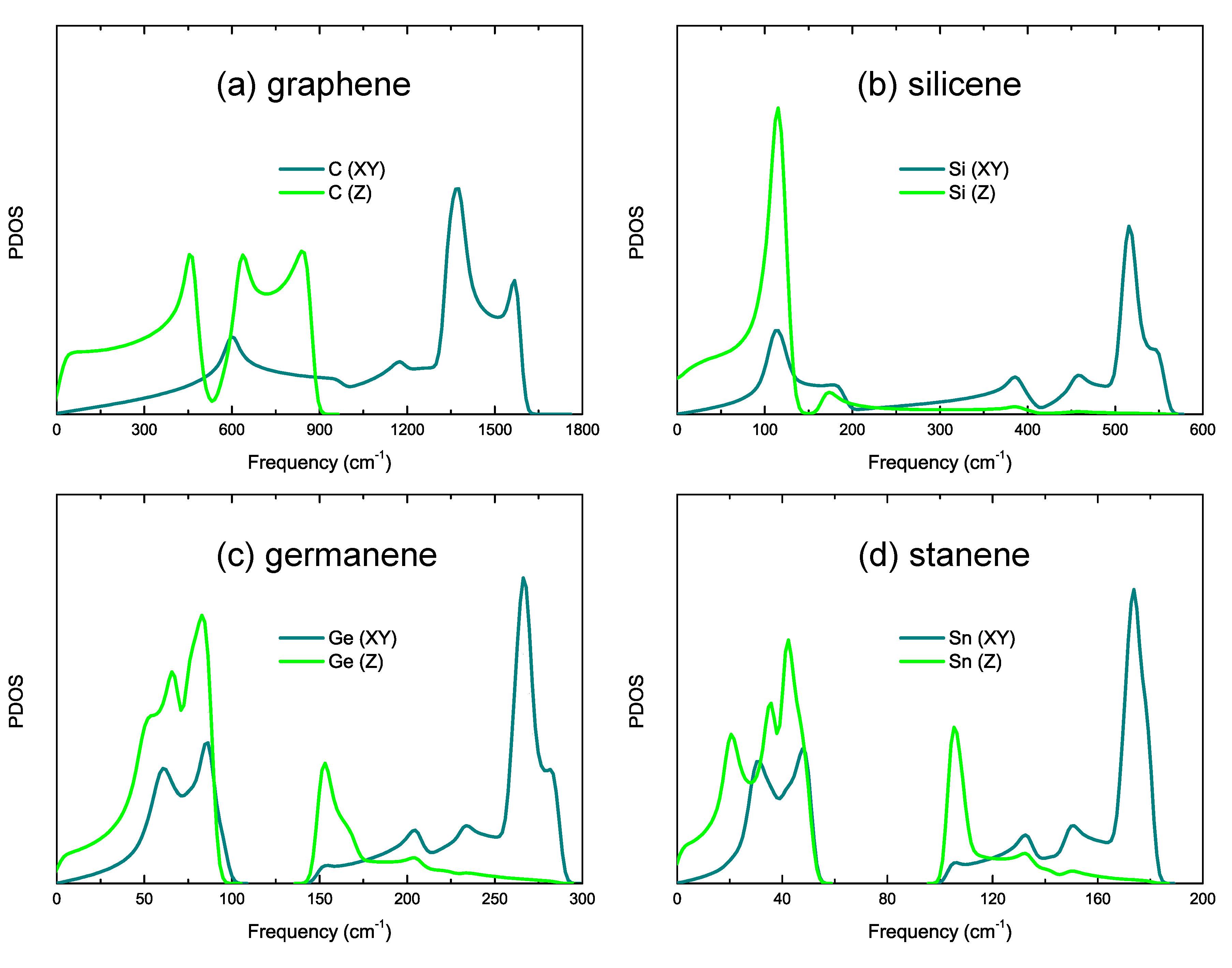}
\caption{Projected PDOS for the XY and Z vibrations of (a) graphene, (b) silicene, (c) germanene and (d) stanene.}
\label{pdos} 
\end{figure*}

Now that we understand that the unexpected higher lattice thermal conductivity of stanene is because the large a-o gap freezes out all $a+a\leftrightarrow o$ processes, and the acoustic bunching of weakens the $a+a\leftrightarrow a$ scattering. Both scattering mechanisms are related to the curvature of phonon dispersion. Thus it is still worthy to investigate the vibrational origin of these differences, especially for the formation of the a-o gap. In the diatomic linear chain model where the scale of the acoustic (optic) phonon branch is governed by the larger (smaller) mass, the formation of the a-o gap is due to the mass difference. However, in contrast to the diatomic linear chain model, only monatomic honeycomb structures are considered herein. To understand the vibrational origin of the a-o gap, we examine the corresponding projected phonon density of states (PDOS) for the in-plane (XY) and out-of-plane (Z) vibrations of all studied 2D group-IV crystals in Fig.~\ref{pdos}.

There is no a-o gap for graphene and silicene, while germanene and stanene show an obvious a-o gap. The formation of the a-o gap is attributed to the vibrational hybridization in the buckled structure. For planar graphene, the XY modes (TA, LA, TO, and LO) are completely decoupled with the Z modes (ZA and ZO), while for other 2D group-IV materials the covalent bonds become nonorthogonal to both the XY plane and Z direction due to the buckled structure, which results in the hybridization between the vibrations in the XY plane and Z direction \cite{Huang2014,Huang2015,Huang2015a}. Analysis of the harmonic IFCs shows that the XZ and YZ components of silicene, germanene and stanene are non-zero, while zero for graphene. If we set the off-diagonal components XZ and YZ of the IFCs to be zero, the gap will disappear due to the lack of the LA-ZO coupling. The LA-ZO coupling has two main effects on the phonon dispersion: (i) flattening the LA branch and (ii) stiffening the ZO branch, leading to the formation of the a-o gap (for details, see Fig.~S2 in the supplemental material). It should be also noticed that the TA branch (highest acoustic phonon branch at the K point) remains the same with or without the LA-ZO coupling, which is similar to the TO branch.

It has been demonstrated recently that buckled structures usually lead to lower thermal conductivity due to the breaking of the out-of-plane symmetry \cite{Lindsay2010,Jain2015}. Here we show that buckled structures also result in a large a-o gap, which tends to reduce the thermal resistance. Thus, when estimate the lattice thermal conductivity in a buckled system, the competing effects between the symmetry selection rule of $a+a\leftrightarrow a$ scattering and the optical scattering channels for $a+a\leftrightarrow o$ scattering should both be taken into account.

\begin{figure}
\centering
\includegraphics[width=0.9\linewidth]{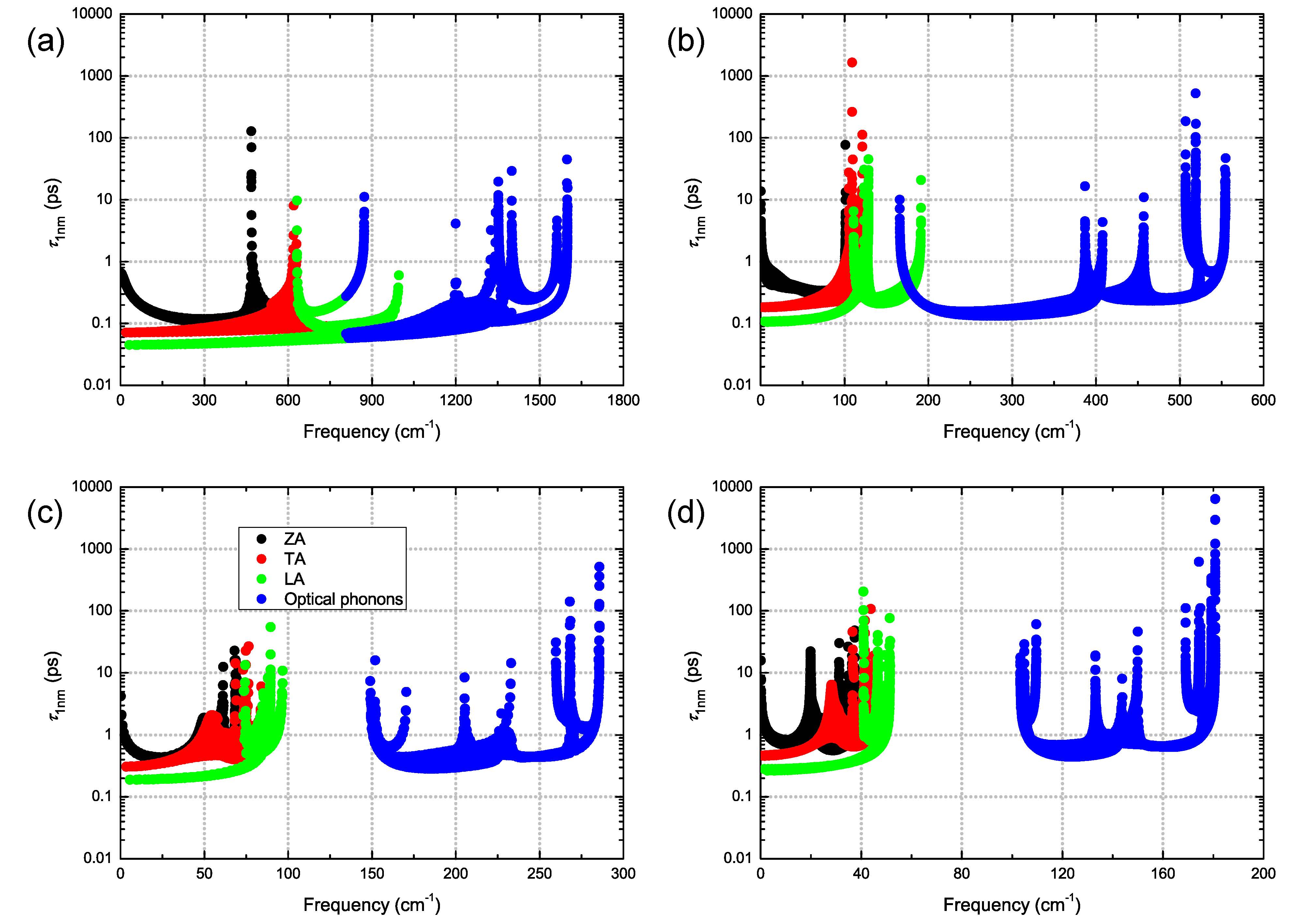}
\caption{Boundary scattering relaxation time of (a) graphene, (b) silicene, (c) germanene and (d) stanene at 300 K for a characteristic length $L = 1$ nm and a specularity parameter $p = 0$.}
\label{boundary} 
\end{figure}

\subsection{Size dependence of $\kappa$}

Silicene, germanene and stanene are expected to realize higher TE efficiency because of much lower lattice thermal conductivity compared to graphene. This will motivate a systematic examination of the width dependence of thermal conductivity in these materials. When the size goes down to nanoscale, the boundary scattering plays an important role in thermal transport. In the Casimir's classical limit, the phonon mean free paths (MFPs) in thin wires under diffuse boundary conditions yield \cite{Casimir1938,Carrete2011}
\begin{equation}
L_{\textrm{c}}=\frac{1+p}{1-p}D,
\end{equation}
where $p$ is the coefficient of specularity, ranging from 0 for diffusive scattering to 1 for mirrorlike reflection considering the effect of partly specular surface, and $D$ is the wire diameter. Here we present the boundary scattering relaxation time for a characteristic length $L = 1$ nm and a specularity parameter $p = 0$ in Fig.~\ref{boundary}. 

The longest boundary scattering relaxation times coincide with the peaks of phonon DOS in Fig.~\ref{pdos}. The peaks in phonon DOS correspond to the flattened dispersions in phonon spectrum, leading to lower group velocities in Fig.~\ref{vg}, and with the same boundary scattering limited MFP $L = 1$ nm, lower group velocities mean longer relaxation time. It should be noticed that phonons with small group velocities are not effective carriers of heat \cite{Feldman1993}. Thus orders of magnitude reduction of $\kappa$ is expected in nanostructured 2D group-IV materials.

Though Casimir's diffuse boundary MFP sheds light on the thermal transport in nanoscale, it should be noticed that the Matthiessen's rule $1/L=1/L_{\textrm{c}}+1/L_{\textrm{intrinsic}}$ cannot be defined for 2D ribbons, because for 3D thin film the MFP is strongly dominated by the phonon scattering on the top and bottom surfaces, while this scattering mechanism does not exist in 2D systems \cite{Wang2011}. Thus the Matthiessen approach underestimates the phonon conductance in 2D nanoribbons, and a complete solution of the BTE is required \cite{Wang2011,ShengBTE}.

\begin{figure}
\centering
\includegraphics[width=0.9\linewidth]{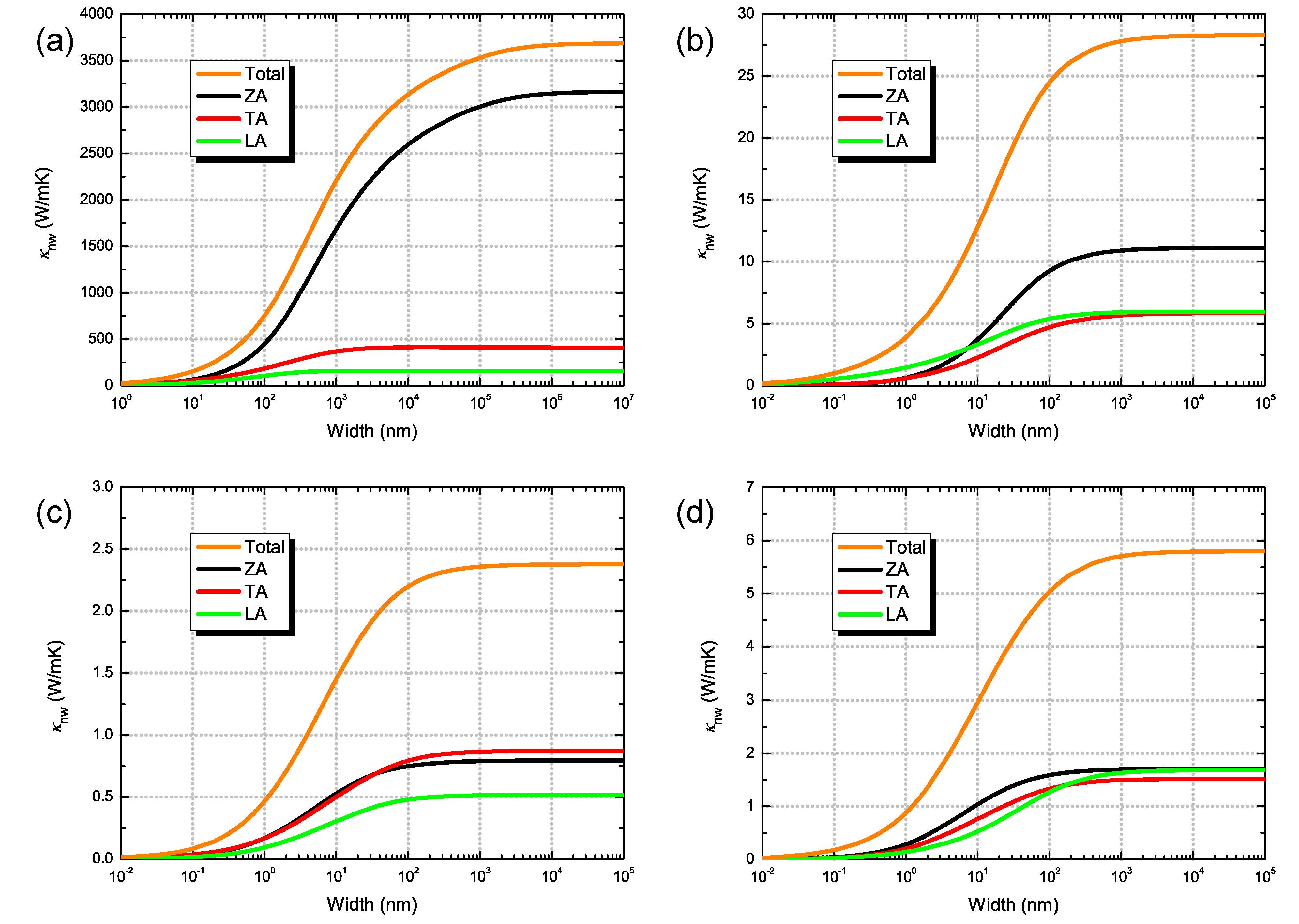}
\caption{Thermal conductivities of (a) graphene, (b) silicene, (c) germanene and (d) stanene nanoribbons at 300 K along [100] direction as a function of width.}
\label{size} 
\end{figure}

Fig.~\ref{size} presents the thermal conductivities of these nanoribbons as a function of width at 300 K along [100] direction. In 2D nanoribbons, phonons with long MFPs will be strongly scattered by the boundary, and the contribution of these phonons to $\kappa$ will be limited. As mentioned above, the ZA contribution decreases from graphene to silicene to germanene and to stanene, therefore the decrease in the ZA contributed $\kappa$ becomes less sensitive to decreasing width. For graphene, the large ZA contribution to $\kappa$ comes from phonons with MFPs below 1 mm, thus the lattice thermal conductivity is very sensitive to boundary scattering, and can be easily reduced in nanostructures for engineering thermal transport. For silicene, germanene and stanene, the ZA contribution to $\kappa$ comes from phonons with MFPs below 1 $\mu$m, and the $\kappa$ is much less sensitive to nanostructuring size effects. Lower lattice thermal conductivity arises typically due to stronger intrinsic phonon scattering, which causes phonons to have shorter MFPs \cite{Li2012}. Consequently, these short-MFP phonons are hard to block by boundaries of a similar size. Thus, it is hard to reduce $\kappa$ further in silicene, germanene and stanene nanostructures. 

\section{Conclusion}

We calculate the lattice thermal conductivity $\kappa$ of 2D group-IV materials using first-principles calculations and an iterative solution of the BTE for phonons. The $\kappa$ decreases monotonically with decreasing Debye temperature from graphene, silicene to germanene, which can be explained by conventional Slack's theory. However, unconventional behavior of $\kappa$ is observed in stanene. Compared to germanene, a large a-o gap in stanene cancels the scattering between acoustic and optical phonons, and a bunching together of the acoustic phonon branches weakens scattering between acoustic phonons, leading to higher lattice thermal conductivity. Though it has been demonstrated that in a buckled system, the thermal conductivity is reduced due to the break of the symmetry selection rule, which severely suppresses the contribution from ZA phonons. Here we show that the buckled structures also induce the LA-ZO coupling, which enlarges the a-o gap and subsequently results in higher lattice thermal conductivity. Thus in a buckled system, the competing effect between $a+a\leftrightarrow a$ scattering and $a+a\leftrightarrow o$ scattering is need to take into account. We also investigate the boundary scattering relaxation time for a characteristic length $L = 1$ nm and a specularity parameter $p = 0$. In 2D nanoribbons, it is hard to reduce $\kappa$ further in silicene, germanene and stanene nanostructures since short-MFP phonons are hard to block by boundaries of a similar size.

\section*{Acknowledgement}
We acknowledge kind help from Dr. Jes\'us Carrete and Prof. Natalio Mingo from CEA-Grenoble. This work is supported by the National Natural Science Foundation of China under Grants No. 11374063 and 11404348, and the National Basic Research Program of China (973 Program) under Grant No. 2013CBA01505.


%

\end{document}